\documentstyle[12pt,twoside,fleqn,espcrc1]{article}
\newcommand{\be}{\begin{equation}}
\newcommand{\ee}{\end{equation}}
                                  
\newcommand{\AmS}{{\protect\the\textfont2                                      
  A\kern-.1667em\lower.5ex\hbox{M}\kern-.125emS}} 
\def\la{\mathrel{\mathpalette\fun <}}
\def\ga{\mathrel{\mathpalette\fun >}}
\def\fun#1#2{\lower3.6pt\vbox{\baselineskip0pt\lineskip.9pt
\ialign{$\mathsurround=0pt#1\hfil##\hfil$\crcr#2\crcr\sim\crcr}}}
\title{In-medium modification of $\rho$-mesons
produced in heavy ion collisions.}

\author{V. L. Eletsky,\address{Institut f\"ur Theoretische Physik III,\\ 
       Universit\"at Erlangen-N\"urnberg, D-91058 Erlangen, Germany}%
       \thanks{On leave of absence from: Institute of Theoretical and 
       Experimental Physics, B.Cheremushkinskaya 25, Moscow 117218, Russia}
B. L. Ioffe \address{Institute of Theoretical and Experimental Physics, \\
       B.Cheremushkinskaya 25, Moscow 117218, Russia}
and J. I. Kapusta \address{School of Physics and Astronomy,\\ 
       University of Minnesota, Minneapolis, MN 55455, USA }}    

\begin{document}
\maketitle 

\begin{abstract}
The mass shift $\Delta m_{\rho}$ and width broadening
$\Delta\Gamma_{\rho}$ of $\rho$ mesons produced in heavy ion collisions is
estimated using a general formula which relates the in-medium mass shift of
a particle to the real part of the forward scattering amplitude ${\rm Re}
f(E)$ of this particle on constituents of the medium and $\Delta\Gamma$ to
the corresponding cross section.
It is found  that in
high energy ($E/A \ga 100~$GeV) heavy ion collisions the $\rho$
width broadening is large and $\rho$ (or $\omega$) peak could hardly be
observed in $e^+e^-(\mu^+\mu^-)$ effective mass distributions.  In low
energy collisions ($E/A \sim$ a few GeV) a broad (a few hundred MeV)
enhancement is expected at the  position of the $\rho$ peak.
\end{abstract}

\section{INTRODUCTION}

The problem of how the properties of hadrons change in hadronic or 
nuclear matter
in comparison to their free values has attracted a lot of attention.
Among these properties of immediate interest are the
in-medium particle's mass shift and width broadening. 
Different models as well as "model independent"
approaches were used to calculate these effects both at finite
temperature and finite density (for a review, see e.g. \cite{ab}).  It is
clear on physical grounds that the in-medium mass shift and width
broadening of a particle are only due to its interaction with the
constituents of the medium. Thus one can use phenomenological information
on this interaction to calculate the mass shift and width broadening. 
In a recent paper\cite{ei} two of us argued that the mass
shift of a particle in medium can be related to the forward scattering
amplitude $f(E)$ of this particle on the constituents of the medium 
(in medium rest frame)

\be
\Delta m (E) = -2\pi\frac{\rho}{m}{\rm Re} f(E) \, .
\label{dm}
\ee
Here $m$ is the vacuum mass of the particle, $E$ is its energy in the rest
frame of the constituent particle, and $\rho$ is the density of consituents.
The normalization of the amplitude corresponds to the standard form of the
optical theorem

\be
k\sigma = 4\pi {\rm Im} f(E) \, ,
\ee
\label{opt}
where $k$ is the particle momentum. The width broadening is given by

\be
\Delta \Gamma(E) = \frac{\rho}{m}~k\sigma(E) \, .
\ee
The applicability criteria of Eqs. (1) and (3) were discussed in [2].
In short, they are:

1) The particle wave length $\lambda$ must be much less than the mean
distance between medium constituents $d$, $\lambda=k^{-1}\ll d$. This means
that the particle momentum $k$ must be larger than a few hundred MeV;

2) The particle formation length $l_f\sim (E/m)/m_{char}$ $(m_{char}\approx
m_{\rho})$ must be less than the nucleus radius $R$;

3) ${\rm Re} f(E)$  which enters Eq.(1) must satisfy the inequality 
$\mid {\rm Re} f \mid < d$;

4) The scattering takes place mostly at small angles, $\theta \ll 1$. 
Only in this case the optical analogy on which Eqs. (1) and 
(3) are based is correct.

Eqs. (1) and (3) are correct also in cases when the medium constituents
have some momentum distributions, e.g. the Fermi distributions for nucleons, 
or represent a gas at a finite temperature. In such cases, in the right-hand
sides of Eqs. (1) and (3) an averaging over the momentum 
distribution of constituents must be performed. Eqs. (1) and (3) were derived 
in [2] basing on simple quantum
mechanical arguments and the optical analogy. This approach allowes one to
formulate explicitly the applicability conditions presented above.
When the medium is a gas in thermal equilibrium the equivalents of Eqs.(1) and
(3) can be derived basing on thermal field theory.  References
\cite{3,4} give  few examples and the reference \cite{5} gives a relativistic
field-theoretic derivation.

In most of the papers on the in-medium hadron mass shifts the hadrons
were considered at rest.  As seen from Eq. (1) this restriction is not
necessary theoretically.  It is  desirable to have theoretical predictions
in a broad  energy interval, since it extends  possibilities of
experimental investigations. As discussed in \cite{6} for the cases of
$\rho$ or $\pi$-mesons embedded in nuclear matter the energy dependence of
the mass shifts is rather significant at low energies, i.e. in the resonance
region.

We  estimate the $\rho$ meson mass shift
and width broadening in the
case of $\rho$-mesons produced in heavy ion collisions.
The  most interesting case is the case of $\rho^0$, which can be observed
through the decay $\rho^0 \to e^+e^-$ $(\mu^+\mu^-)$. We will assume that
$\rho$-mesons are formed at the latest stage of the evolution of hadronic
matter created in the course of a heavy ion collision when the matter can be
considered as a noninteracting gas of pions and nucleons. (We will neglect
the admixture of kaons and hyperons, which is known to be small \cite{7}, as
well as heavy resonances.) This stage is formed when, during the evolution
of the matter after collision, the total density of nucleons and pions becomes
of order of the normal nucleon density in nuclei. The description of nuclear
matter as a noninteracting gas of nucleons and pions of course cannot be
considered as a very good one. So, it is clear from the beginning that our
results may be only semiquantitative. The main ingredients of our calculation
are the $\rho\pi$ and $\rho N$ forward scattering amplitudes and total cross
sections as well as the values of nucleon and pion densities.

We consider here central heavy ion collisions and assume that the
nucleon and pion momentum distributions in the gas are just the momentum
distributions measured experimentally in such collisions. (Another model,
in which nucleons and pions are assumed to be in the state of ideal gas at
fixed temperature and chemical potential, will be considered in a subsequent
publication [8].)

\vspace{5mm}

\noindent
\section{CALCULATION OF $\rho \pi$ AND $\rho N$ CROSS SECTIONS
AND FORWARD SCATTERING AMPLITUDES.}

\vspace{3mm}

Let us first focus on the amplitudes and cross sections. To determine
these quantities we use the following procedure. In the low energy region we
saturate the cross sections and forward scattering amplitudes by resonance
contributions. At high energies we determine $\sigma_{\rho N}$  and
$\sigma_{\rho \pi}$  from $\sigma_{\gamma N}$  and $\sigma_{\gamma\pi}$
using vector dominance model
(VDM). $\sigma_{\gamma N}$  is well known experimentally \cite{9},
${\rm Re}f_{\gamma N}$ is determined from dispersion  relation; 
$\sigma_{\gamma
\pi}$ and  ${\rm Re}f_{\gamma \pi}$ can be obtained by the Regge approach.
Since VDM allows one to find only the cross sections
of transversally polarized $\rho$-mesons, we restrict ourselves to this
case.  As was shown in [2], at $E_{\rho}\ga 2$~GeV $\Delta m$ and $\Delta
\Gamma$ for longitudinal $\rho$-mesons in nuclear matter are much smaller
than for transverse ones.  At zero $\rho$-meson energy $\Delta m$ and
$\Delta \Gamma$  for transverse and longitudinal $\rho$-mesons are evidently
equal. In the case of scattering on low temperature gas they are comparable
[10]. Therefore, for unpolarized $\rho$-meson our results should be 
multiplied by a factor ranging from 1 to 3/2.

To estimate ${\rm Re} f_{\rho\pi}(s)$ at low energy we write in the center
of mass (c.m.) frame

\be
{\rm Re} f_{\rho\pi}(s)=-\sum_{R} F_s F_i \frac{1}{2q_{cm}}
\frac{B_R\Gamma_R (\sqrt{s}-m_R)}{(\sqrt{s}-m_R)^2 +\Gamma_R^2/4} \, ,
\ee
\label{res}
where $\sqrt{s}$ is the total c.m. energy, $m_R$ and $\Gamma_R$ are the mass
and the total width of the resonance, $B_R$ is the branching ratio of its decay
into $\pi\rho$ and $q_{cm}$ is the center of mass momentum,

\be
q_{cm} =\frac{1}{2\sqrt{s}}\sqrt{[s-(m_{\rho}+m_{\pi})^2]
[s-(m_{\rho}-m_{\pi})^2]} \, .
\ee
\label{qcm}
The factor $F_s$ is the spin factor,
$F_i$ is the isospin factor. The latter is equal to $1/3=(1/2)\times (2/3)$
for $I_R=1$. The first factor here reflects the fact that we are interested
in the $\rho^0 \pi$ scattering and only one of the two decay channels of a 
$I_R =1$ resonance can contribute here: $R^{\pm} \to \rho^0 \pi^{\pm}$, but 
not $\rho^{\pm} \pi^0$. The second factor corresponds to the assumption, that
all three pion isospin  states are equally populated in the gas.
Similarly, for $I_R=0$ the isospin factor is $1/9=(1/3)\times (1/3)$.  We
take into account the following resonances [9]:  $a_1(1260)$,
$\pi(1300)$, $a_2(1320)$ and $\omega(1420)$. The nearest
resonance under the threshold, $\omega(782)$, does not give a considerable
contribution due to its narrow width.
For the spin factor we take $F_s= 1,~1,~2,~1$
correspondingly for the above mentioned resonances. (These factors take into
account that we consider only transvers $\rho$). The amplitude in the
pion rest frame is obtained from Eq. (4) multiplying it by the rescaling
factor $k_{\rho}/q_{cm} = \sqrt{s}/m_{\pi}$, where $k_{\rho}=
\sqrt{E_{\rho}^2-m_{\rho}^2}$ is the $\rho$ momentum in the pion rest frame.

For $\sigma_{\rho\pi}$ we use the standard resonance formula

\be
\sigma_{\rho \pi} =
\sum_R~F_sF_i\frac{\pi}{q^2_{cm}}\frac{B_R\Gamma^2_R}{(\sqrt{s} -m)^2 +
\Gamma^2_R/4} \, .
\ee

As is known, the pion scattering amplitude on any hadronic target vanishes at
zero pion energy in the target rest frame in the limit of massless pions
(Adler theorem). In the framework of effective Lagrangians this can be
achieved if the pion field enters through its derivative,
$\partial  \varphi/\partial x_{\mu}$. We assume that in the interaction 
describing the $\rho\pi$
scattering through the $a_1$ resonance, $\partial  \varphi/\partial x_{\mu}$ 
is multiplied by the $\rho$-meson field strength tensor $F_{\mu\nu}$ and 
the field $a_{1\nu}$. This results in appearance of an additional factor in 
$a_1$ contributions to ${\rm Re}f_{\rho\pi}$ and $\sigma_{\rho\pi}$ in 
Eqs. (4) and (6)

\be
\Biggl ( \frac{s - m^2_{\rho} - m^2_{\pi}}{m^2_{a_1} - m^2_{\rho} -
m^2_{\pi}}\Biggr )^2 \, ,
\ee
which is normalized to 1 at $s=m^2_{a_1}$. At $s > m^2_{a_1}$
this factor was not taken into account. Similar factors were also
introduced for contributions of other resonances.

In the high energy region we assume that the Regge approach is valid for 
$\gamma\pi$  scattering and apply VDM to relate $\rho
\pi$ and $\gamma \pi$ amplitudes. As is well known, the Regge pole
contributions to the forward scattering amplitude normalized according to
Eq. (2) have the form

\be
f(s) = -\frac{k}{4\pi s}\sum_i
~\frac{1+e^{-i\pi\alpha_i}}{\sin\pi\alpha_i}s^{\alpha_{i}}r_i \, ,
\ee
where $\alpha_i$  is the intercept of $i$-th  Regge pole trajectory,
$r_i$ is its residue, $k$ is the projectile momentum in the target rest
frame. As follows from Eqs. (2) and (8),

\be
\sigma(s) = \sum_i ~r_i s^{\alpha_i -1} \, ,
\ee

\be
{\rm Re} f(s) = -\frac{k}{4\pi s}\sum_i~\frac{1+\cos\pi \alpha_i}{\sin\pi
\alpha_i}r_is^{\alpha_i} \, .
\ee
For $\sigma_{\gamma \pi}$  only the $P$ (pomeron) and $P^{\prime}$ poles
contribute [11,12]. The residues of $P$ and $P^{\prime}$ poles in
$\gamma\pi$  scattering were found by Boreskov, Kaidalov and Ponomarev (BKP)
[12] using the Regge pole factorization theorem and the data on $\gamma p$,
$\pi p$  and $pp$ scattering. Taking the BKP values for the $P$ and 
$P^{\prime}$ residues we have

\be
\sigma_{\pi \gamma}(s) = 7.48 \alpha ~\Biggl [
\Biggl (\frac{s}{s_0}\Biggr  )^{\alpha_P-1} + 0.971
\Biggl (\frac{s}{s_0}\Biggr  )^{\alpha_{P^{\prime}}-1} \Biggr ] \, ,
\ee
where $\alpha_P=1.0808,~\alpha_{P^{\prime}}=0.5475$, $\alpha=1/137$,
$s_0=1$ GeV$^2$ and $\sigma$ in Eq. (11) is given in millibarns, if $s$ is 
in GeV$^2$. For the $P$ and $P^{\prime}$
intercepts we take the Donnachie-Landshoff values [13]. Since in their fit to
the data BKP assumed $\alpha_P=1$ and $\alpha_{P^{\prime}}=1/2$, the values of
residues in Eq. (11) are slightly changed in comparison with [12] in order to
get the same $\sigma_{\pi\gamma}$ at  $s=9$~GeV$^2$. From Eqs. (10) and (11) 
the real part of the forward $\gamma \pi$ scattering amplitude can be found:

\be
{\rm Re}f_{\gamma\pi}(s)_{\pi~rest~frame} = -\frac{k}{4\pi} 7.48\alpha \left \{
-0.106\Biggl ( \frac{s}{s_0}\Biggr )^{\alpha_P-1} + 0.752
\Biggl ( \frac{s}{s_0}\Biggr )^{\alpha_{P^{\prime}}-1}\right \} \, ,
\ee
where the momentum $k$ is in GeV and ${\rm Re}f$ is in mb$\cdot$GeV.

In VDM  $\sigma_{\rho\pi}(s)$ is related to $\sigma_{\gamma \pi}(s)$ by (see
[2])

\be
\sigma_{\rho \pi}(s) = \frac{g^2_{\rho}}{4\pi \alpha} \Biggl (1 +
\frac{g^2_{\rho}}{g^2_{\omega}} \Biggr )^{-1} \sigma_{\gamma \pi}(s) \, ,
\ee
where $g^2_{\rho}/4\pi =2.54$, $g^2_{\rho}/g^2_{\omega}=1/8$ and the
$\varphi$-meson contribution is neglected. A similar relation holds
also for ${\rm Re}f_{\rho\pi}$. Unlike in [2], we prefer here to use 
direct Regge formulae for ${\rm Re}f$ at high energies instead of finding 
it from $\sigma$ through the dispersion relation, since in the latter
approach the results are sensitive  
to the low energy domain which is more uncertain.

The  results of the calculations of $\sigma_{\rho\pi}$ and 
${\rm Re}f_{\rho\pi}$ as
functions of $\rho$-meson energy in the pion rest frame are presented in Fig.1.
As seen from Fig.1, the matching of low and high energy curves is
satisfactory.

For the amplitude ${\rm Re} f_{\rho N}$ at laboratory frame energies of $\rho$
above 2 GeV we use the results of Ref.\cite{ei} obtained using dispersion
relation, VDM and experimental data on $\sigma_{\gamma N}$. At lower
$E_{\rho}$ we again use the resonance approximation

\be
{\rm Re} f_{\rho N}(s)=-\frac{1}{4}~ \frac{1}{2q_{cm}}\sum_{R}(2J_R+1)
F_i
\frac{\Gamma_R^{\rho N} (\sqrt{s}-m_R)}{(\sqrt{s}-m_R)^2 +\Gamma_R^2/4}
\ee
(the factor $(1/4)$ appears because we consider only  transverse $\rho$).
The isospin  factors $F_i$ are 1/3 and 2/3 correspondingly for $N$  and
$\Delta$ resonances.
We take 10 well established $N$ and $\Delta$ resonances[9] with significant
branchings into $\rho N$ and with masses above the $\rho N$ threshold and 
below 2200 MeV.
The set of baryonic resonances taken into account is close to the set used in
[14]. The main difference in comparison with [14] is that 
effective widths $\Gamma_{eff}^{\rho N}=
\Gamma_R^{\rho N}(q_{cm}/q_{cm,R})^{2l+1}$ were introduced only for the
resonances close to the $\rho N$ threshold ($q_{cm,R}$ is the value of 
$q_{cm}$ at the resonance). At $q_{cm}> q_{cm R}$ we put
$\Gamma_{eff}^{\rho N}=\Gamma_R^{\rho N}$.
Apart from these resonances, in the calculation of ${\rm Re}f_{\rho N}$ two
resonances with masses below the $\rho N$ threshold were also accounted for:
$\Delta(1238)$ and $N(1500)$. It was assumed that VDM relates 
$\Gamma_{\rho N}$ and $\Gamma_{\gamma N}$ of these resonances in the
following way. Since both resonances are close to the $\rho N$ threshold, we
can write for each of them: $\Gamma_{\rho N}=q_{cm}\gamma_{\rho N}$,
$\Gamma_{\gamma N}=k_{cm}\gamma_{\gamma N}$, where $q_{cm}$  and $k_{cm}$
are the $\rho N$ and $\gamma N$ c.m. momenta, respectively. Then we assume
that $\gamma_{\rho N}$ and $\gamma_{\gamma N}$ are related by the VDM formula

\be
\gamma_{\gamma N} = 4\pi \alpha \frac{1}{g^2_{\rho}}\Biggl ( 1 +
\frac{g^2_{\rho}}{g^2_{\omega}} \Biggr ) \gamma_{\rho N} \, .
\ee
The values of $\gamma_{\gamma N}$ can be found from the values of
$\sigma_{\gamma N}$ at the resonance peaks. The contributions of
$\Delta(1238)$  and $N(1500)$ to ${\rm Re}f_{\rho N}$  are essential at low
energies of $\rho$ mesons: they comprise about $(-1) - (-0.5)$ fm  at
$E_{\rho}=1 - 2$ GeV in the nucleon rest frame.

The results for
$\sigma_{\rho N}$ and ${\rm Re} f_{\rho N}$ in the rest frame of the nucleon
together with the curve obtained in Ref.\cite{ei} for high energies and the
matching curve are shown in Fig.2. As can be seen, the matching of low
energy and high energy curves is good.

\vspace{90mm}
\vspace{5mm}
\noindent
\section{DETERMINATION OF THE MASS SHIFT AND WIDTH BROADENING
OF THE $\rho$-MESON PRODUCED IN HEAVY ION COLLISION
FROM THE DATA ON NUCLEON AND PION DISTRIBUTIONS}

\vspace{3mm}
As was mentioned above, in heavy ion collisions we will consider only 
nucleons and pions as constituents of the medium. Therefore, in this case
Eqs. (1) and (3) take the form

\be
\Delta m_{\rho}(E) = -\frac{2\pi}{m}\left \{ \rho_N{\rm Re}f_{\rho N}(E)+
\rho_{\pi}{\rm Re}f_{\rho\pi}(E)\right \} \, ,
\ee

\be
\Delta\Gamma_{\rho}(E) = \frac{k}{m}\left \{ \rho_N~
\sigma_{\rho N}(E)+\rho_{\pi}\sigma_{\rho\pi}(E)\right \} \, ,
\ee
where $\rho_N$  and $\rho_{\pi}$ are the nucleon  and pion densities in the
medium formed at the latest stage of evolution of hadronic matter produced
in heavy ion collisions.

We will restrict ourselves to consideration of head-on (central) collision
with small values of impact parameter when the number of participants --
the nucleons which experience considerable momentum transfer -- is close to
the total number of colliding nucleons.

Experimental data shows that the nucleons and pions produced in
heavy ion collisions cannot be considered as thermal gas even at the final
stage of evolution of hadronic matter created in the collisions.
In order to demonstate this
let us discuss separately the cases of high energy ($E/A \sim 100$ GeV)
and low energy ($E/A \sim$ a few GeV) heavy ion collisions. In case of
high energy collisions the longitudinal (relative to the beam) and
transverse momenta of  nucleons and pions are very different. In the
experiment on S + S  collisions at $E/A=200$ GeV [15]  it was
found (in the centre of mass frame) that $\langle p^{cm}_{LN}\rangle =
3.3$ GeV, $\langle p_{TN}\rangle = 0.61$ GeV and 
$\langle p^{cm}_{L\pi}\rangle
\approx 0.70$ GeV, $\langle p_{T\pi}\rangle
\approx 0.36$ GeV. In other experiments on high energy heavy ion collisions
(see e.g. [16,17]) the situation is qualitatively similar.
This means that in this case one can by no means speak about thermal gas
of final particles and their momentum distributions must be taken from
experiment.

The data on low energy heavy ion collisions also indicate that pions and
nucleons cannot be described as gases in thermal equilibrium.
The  angular distributions of pions produced in Ni + Ni  collisions at
$E/A=(1-2)$ GeV show a considerable anisotropy[18]. If the pion
angular distribution in the centre of mass frame is approximated by $1+
a\cos^2 \theta$, then it follows from the data that $a \approx 1.3$. 
Unfortunately, there is not enough information in the data about the nucleon 
angular and momentum  distributions. We have checked  the hypothesis of thermal
equilibrium by assuming that the probability  of production of a given
number of particles is proportional to the statistical weight of the final
state (Fermi-Pomeranchuk approach [19,20]). It is evident that this
hypothesis is even more general  than the hypothesis of thermal
equilibrium.  In this approach the pion/nucleon ratio $R_{\pi}=n_{\pi}/N$ in
central collisions can be predicted in terms of the main ingredient of the
method -- the volume per particle at the final stage of evolution (and, of
course, the initial energy). The calculations show that the data [18]
on the energy dependence of the ratio $R_{\pi}$ are well described 
by the statistical model, but in order to get the absolute values of
$R_{\pi}$ in Ni + Ni  as well as in Au + Au [21] collisions, it is
necessary to put the volume per nucleon very small, about 5 times smaller,
than the one in a normal nucleus, which is unacceptable.

Therefore, the only way to perform the averaging over momentum distributions
of pions and nucleons seems to take the latter from experimental data on heavy
ion collisions.

In calculation of the $\rho$-meson mass shift and
width broadening the averaging must be performed over the $\rho$-meson
direction of flight relative to nucleons and pions. Such calculation can
be done only for real experimental conditions. For this reason we restrict
ourselves to crude estimates.

Consider first the case of high energies  and take as an example the
experiment [15] on central collisions, where the ratio of
pion to nucleon multiplicities was found to be $R_{\pi} =5.3$ .  Suppose
that in this experiment the $\rho$-meson is produced with longitudinal and
transverse momenta in the lab. frame $k_{L\rho}=3$ GeV,
$k_{T\rho}=0.5$ GeV. We choose these values as typical for such experiment.
At such values of $\rho$ momenta the formation time of $\rho$-meson is close
to the mean formation time of pions produced in heavy ion collisions.
So, a necessary condition of our approach is fulfilled. Since the mean
momenta of nucleons and pions in the experiment [15] are known (they
were presented above), it is possible using the curves of Figs. 1 and 2
to calculate the mean values of ${\rm Re}f_{\rho N}$, ${\rm Re}f_{\rho\pi}$,
$\sigma_{\rho N}$ and $\sigma_{\rho\pi}$ in $\rho N$ and $\rho \pi$
scattering. The results are (in the lab. frame):

\be
\langle{\rm Re}f_{\rho N}\rangle \approx -1.1~fm \, , ~~~~~~~
\langle{\rm Re}f_{\rho\pi}\rangle \approx 0.03~fm \, ,
\ee

\be
\langle\sigma_{\rho N}\rangle \approx 43 ~mb \, ,~~~~~~~
\langle\sigma_{\rho\pi}\rangle\approx 20~mb \, .
\ee
The small value of $\langle{\rm Re}f_{\rho\pi}\rangle$ comes from the 
compensation of positive and negative ${\rm Re}f_{\rho\pi}$ at low and 
high energies (Fig. 1), i.e. from the scattering of $\rho$ on pions moving in 
the same direction as
$\rho$ (comovers), or in the opposite one. Because of this compensation
$\langle {\rm Re}f_{\rho\pi}\rangle$ is badly determined, but since 
it is small this does not affect the final result.
Using Eqs. (16) and (17) we can now find the mass shift and
width broadening of $\rho$-meson. For the nucleon and pion densities we take

\be
\rho_N = \frac{N}{V} = \frac{N}{Nv_N + nv_{\pi}} = \frac{1}
{v_N(1+R_{\pi}\frac{v_{\pi}}{v_N})} \, ,
\ee

\be
\rho_{\pi} = \frac{n}{V}= \frac{n}{Nv_N + nv_{\pi}} =
R_{\pi} \frac{1}{v_N(1 + R_{\pi}\frac{v_{\pi}}{v_N})} \, ,
\ee
where $N$  and $n$ are the numbers of nucleons and pions at the final stage
of evolution, $R_{\pi}=n/N$, $V$ is the volume of system at this stage. It
is assumed that at this stage any participant (nucleon or
pion) occupies a definite volume $v_N$ or $v_{\pi}$.
We can write

\be
\rho_N=\rho^0_N \frac{1}{1 + R_{\pi}\beta} \, , ~~~~\rho_{\pi} =
\rho^0_NR_{\pi}\frac{1}{1+R_{\pi}\beta} \, ,
\ee
where $\rho^0_N=1/v_N$ and $\beta=v_{\pi}/v_N$.
For numerical estimates we take
$\rho^0_N=0.3$ fm$^{-3}$, almost two times the standard nucleous density. This
number is probably one of the most uncertain ingredients of our
calculations. Substitution of Eqs. (18),(19) and (22) into Eqs. (16) and 
(17) gives (for the experimental value of $R_{\pi} = 5.3$ and $\beta=1$)

\be
\Delta m_{\rho} = 18 - 2 = 16~MeV \, ,
\ee

\be
\Delta \Gamma_{\rho} \approx 150 + 400 = 550~MeV \, .
\ee
The first numbers in Eqs. (23) and (24) refer to the contributions from 
$\rho N$ and
second ones from $\rho\pi$  scattering. Because $\rho$-meson width broadening
appears to be very large, the basic condition of our approach, 
$\Delta \Gamma_{\rho}\ll m_{\rho}$ is fullfilled badly. 
The other applicability condition of
the method, $\mid {\rm Re}f \mid < d$, is also not well satisfied, 
since in this
case $d=0.9$ fm. For these reasons the values of $\Delta m_{\rho}$ in Eq. (23)
and of $\Delta\Gamma_{\rho}$ in Eq. (24) may be considered only as estimates.

The main conclusion from Eqs. (23) and (24) is that in the case of
$\rho$-mesons  produced in high energy heavy ion collisions with the 
longitudinal and transverse momenta chosen
above, the mass shift is small, but the
width broadening is large and hardly a $\rho$ (or $\omega$) peak would be
observed in $e^+e^-$ or $\mu^+\mu^-$  effective mass distributions. Let us
estimate how sensitive are the results to a variation of $k_L$ and $k_T$. It
can be easily seen that the mass shift will be small in all cases (say,
$\Delta m_{\rho} \la 50$ MeV). If we put $k_{T\rho}=0$  instead of
$k_{T\rho}=0.5$ GeV,  this will only weakly affect the mean value of
$\sigma_{\rho N}$ and decrease $\sigma_{\rho\pi}$ by 20\%. The latter
results in decrease of $\Delta \Gamma_{\rho}$ in Eq. (24) by $80$ MeV, 
i.e. within the limits of accuracy. The variation of $k_{\rho L}$ between 
$1$ GeV and $10$ GeV also results in variation of the same order (10-20\%) 
in $\Delta\Gamma_{\rho}$ in Eq. (24).

As was mentioned above, the main uncertainty in our approach comes from the
value of the nucleon density at the final stage of evolution which we put to 
be $\rho^0_N = 0.3$ fm$^{-3}$. If this density at the time of the $\rho$-meson 
formation would be, say, two times smaller, then we would have 
$\Delta\Gamma_{\rho}\sim 250$ MeV and the $\rho$-meson could be observed 
as a broad peak in $e^+e^-$  or $\mu^+\mu^-$ mass spectrum. It should be 
mentioned, however, that the chosen above value of $\beta=v_{\pi}/v_N=1$
is not the only possible, or even the most plausible. If we assume that
$\beta=(r_{\pi}/r_N)^3$, where $r_{\pi}$ and $r_N$  are pion and nucleon
electromagnetic radii, $r_{\pi}=0.66$ fm, $r_N=0.81$ fm, then $\beta\approx
0.55$. This choice of $\beta$ increases $\Delta\Gamma_{\rho}$ by a factor of
1.6.

Our qualitative conclusion is the following. 
In the central heavy ion collisions at
high energies, $E/A\sim 100$ GeV, where a large number of pions per
participating nucleon is produced, the $\rho$- (or $\omega$) peak will be
(if at all) observed in $e^+e^-$ or $\mu^+\mu^-$ effective mass distributions 
only as a very broad enhancement.  Inspite of the fact that
in our approach we used a few assumptions, which may raise some
doubts (hypothesis of noninteracting nucleon and pion matter at the final
stage of evolution, the numerical value of nucleon density, etc.), we believe
that this qualitative conclusion is still intact.

This conclusion
is in qualitative agreement with the measurement of $e^+e^-$  pair
production in heavy ion collisions [22], where no $\rho$ peak was found and
only a smooth $e^+e^-$ mass spectrum between 0 and 1 GeV was observed.
If, however, such a peak will be observed in future experiments,
this would indicate that the hadronic (nucleon and pion) density at the
final stage of evolution, where $\rho$-meson is formed, is low, lower even
than the normal nuclear density.

Recently preliminary data have been presented [22], where in studying the
$e^+e^-$ mass spectrum in Pb - Au collisions at $E/A=$ 160 GeV it was
found that the $\rho$-peak is absent for $k_T(e^+e^-)< 400$ MeV, but reappears
as a broad enhancement for $k_T(e^+e^-)> 400$ MeV. We do not see a
possibility for this in the framework of our approach in the
case of central heavy ion collisions. Moreover, we believe that for central
collisions the absence of $\rho$-peak at low $k_T$ and its reappearance at
higher $k_T$  will be hard to explain in any reasonable model. The only
explanation we see for this effect (if it will be confirmed), is that in this
experiment the peripheral $\rho$-meson production plays  an essential role.
In this case the $\rho$-mesons with higher $k_T$ have a larger probability 
to escape the collision region and decay as free ones.

Let us turn now to the case of low energy heavy ion collisions, $E/A \sim$
a few GeV. Consider as an example the case of heavy ion collisions at
$E_{kin}/A=3$ GeV and production of $\rho$-meson with the energy
$E^{tot}_{\rho}=1.2$ GeV in the forward direction. (This energy of 
$\rho$-meson was chosen, because our approach works better at
higher $E_{\rho}$, and $\rho$ of this energy can be kinematically produced
at such heavy ion energy). The number of pions, produced in
$E_{kin}/A=3$ GeV collisions can be found by extrapolation of the data
[18] on Ni + Ni  collisions.  The data shows that $R_{\pi}$ is with a good
accuracy linear in $\sqrt{s}/2 -m$. We find $R_{\pi}=0.48$. As follows from
the analysis of the data [18] at $E_{kin}/A=1.93$ GeV, the
energies of produced pions are rather small, $E_{\pi}\sim 200-300~MeV$. At
such low energies it is reasonable to suppose for pions  
$<p_{L\pi}>=<p_{\perp\pi}>\approx 0.2~GeV$.  
Assuming (this assumption does not influence
essentially the final results) that the mean transverse momentum of nucleon
participants produced in the collision is the same as at high energy,
$<p_{T N}>=0.61$ GeV [15] (see above), we can construct the momentum 
distributions of nucleons. Then we are in a position to calculate the mean 
values of
${\rm Re}f_{\rho N}$, ${\rm Re}f_{\rho \pi}$, $\sigma_{\rho N}$ and 
$\sigma_{\rho\pi}$ for this case. The results are:

\be
\langle {\rm Re}f_{\rho N} \rangle = -0.54~fm \, , ~~~~~~~~~
\langle {\rm Re}f_{\rho \pi} \rangle = 0.30~fm \, ,
\ee

\be
\langle \sigma_{\rho N} \rangle = 46 ~mb \, ,~~~~~~~~~~
\langle \sigma_{\rho \pi} \rangle = 13~mb \, .
\ee
For the $\rho$ mass shift and width broadening we have (at the same value of
$\rho^0_N$ as before and $\beta=1$):

\be
\Delta m_{\rho} = 37 - 10 = 27~MeV \, ,
\ee

\be
\Delta\Gamma_{\rho} = 250 + 35 = 285~MeV \, .
\ee
(The first numbers in Eqs. (28) and (29) refer to $\rho N$ scattering, 
the second ones to $\rho\pi$). The conclusion is that at low energy heavy ion
collisions $\rho$-peak may be observed in $e^+e^-$ or $\mu^+\mu^-$ effective
mass distribution as a broad enhancement approximately at the position of
$\rho$-mass.

We are indebted to K. Boreskov, A. Kaidalov, G. Brown and A. Sibirtsev for
illuminating discussions.
B.I. thanks A. Smirnitsky and V. Smolyankin for
the help in getting information about experimental data.

This work was supported in part by INTAS Grant 93-0283, CRDF grant RP2-132,
Schweizerischer National Fonds grant 7SUPJ048716 and by the RFBR grant
97-02-16131.
V. L. E. acknowledges support of BMBF, Bonn, Germany.

\end{document}